\documentclass[useAMS,usenatbib]{mn2e}
\usepackage{url,times,graphicx,amsmath,amsfonts,amssymb,aas_macros,color,epsfig,varioref,subfigure,comment,natbib}
\usepackage[normalem]{ulem}

\newcommand{\Tab}[1]{Table~\ref{#1}}
\newcommand{\Sec}[1]{Section~\ref{#1}}

\newcommand{\Fig}[1]{Fig.~\ref{#1}}
\newcommand{\hMpc}{{\ifmmode{h^{-1}{\rm Mpc}}\else{$h^{-1}$Mpc}\fi}}
\newcommand{\hGpc}{{\ifmmode{h^{-1}{\rm Mpc}}\else{$h^{-1}$Gpc}\fi}}
\newcommand{\hkpc}{{\ifmmode{h^{-1}{\rm kpc}}\else{$h^{-1}$kpc}\fi}}
\newcommand{\SimuLN}{{\rm SimuLN256}}
\newcommand{\SimuLNHi}{{\rm SimuLN512}}
\newcommand{\SimuLG}{{\rm SimuLGzoom}}
\newcommand{\wfcr}{{\rm WF/RZA/CR}}
\newcommand{\lgi}{{\rm LG-gen}}
\newcommand{\lgii}{{\rm LG-dyn}}
\newcommand{\lgiii}{{\rm LG-vel}}
\newcommand{\hMsun}{{\ifmmode{h^{-1}{\rm {M_{\odot}}}}\else{$h^{-1}{\rm{M_{\odot}}}$}\fi}}
\newcommand{\Msun}{{\ifmmode{{\rm {M_{\odot}}}}\else{${\rm{M_{\odot}}}$}\fi}}
\def\hMpc{$h^{-1}\,{\rm Mpc}$}
\def\hkpc{$h^{-1}\,{\rm kpc}$}
\def\LCDM{\ensuremath{\Lambda}CDM}

\title
[Constrained Local UniversE Simulations: A Local Group Factory]
{Constrained Local UniversE Simulations: A Local Group Factory}

\author[Edoardo Carlesi]
{Edoardo Carlesi,$^{1}$
\thanks{E-mail: carlesi@phys.huji.ac.il}
Jenny G. Sorce,$^{2}$
Yehuda Hoffman,$^{1}$
Stefan Gottl\"ober,$^{2}$
Gustavo Yepes,$^{3,4}$\and
Noam I. Libeskind,$^{2}$
Sergey V. Pilipenko,$^{5,6}$
Alexander Knebe,$^{3,4}$
H\'el\`ene Courtois,$^{7}$\and
R. Brent Tully,$^{8}$
Matthias Steinmetz$^{2}$
\\
\\
$^{1}$Racah Institute of Physics, 91040 Givat Ram, Jerusalem, Israel\\
$^{2}$Leibniz-Institut f\"ur Astrophysik Potsdam (AIP), An der Sternwarte 16, D-144 Potsdam, Germany\\
$^{3}$Grupo de Astrof\'isica, Departamento de Fisica Teorica, Modulo C-8, Universidad Aut\'onoma de Madrid, Cantoblanco E-280049, Spain\\
$^{4}$Astro-UAM, UAM, Unidad Asociada CSIC\\
$^{5}$Moscow Institute of Physics and Technology, Institutskij per. 9, 141700 Dolgoprudnyj, Russia\\
$^{6}$Astro Space centre of Lebedev Physical Institute of Russian Academy of Sciences, Profsojuznaja st. 84/32, 117997 Moscow, Russia\\
$^{7}$University of Lyon, UCB Lyon 1/CNRS/IN2P3; IPN Lyon, France\\
$^{8}$Institute for Astronomy (IFA), University of Hawaii, 2680 Woodlaum Drive, HI 96822, US}
\begin{document}

\date{Submitted XXXX January XXXX}

\pagerange{\pageref{firstpage}--\pageref{lastpage}} \pubyear{2015}

\maketitle

\label{firstpage}

%%%%%%%%%%%%%%%%%%%%%%%%%%%%%%%%%%%%%%%%%%%%%%%%%%%

\begin{abstract}
Near field cosmology is practiced by studying the Local Group (LG) and its 
neighbourhood.
The present paper describes a framework for simulating the ``near field'' on 
the computer. 
Assuming the \LCDM\ model as a prior and applying the Bayesian tools of the Wiener filter (WF) and constrained 
realizations of Gaussian fields to the Cosmicflows-2 (CF2) survey of peculiar 
velocities, constrained simulations of our cosmic environment are performed.
The aim of these simulations is to reproduce the LG and its local environment.
Our main result is that the LG is likely a robust outcome of the \LCDM\ scenario when subjected to the constraint derived from CF2  data, emerging in an  environment akin to the observed one.
Three levels of  criteria are used to define the simulated LGs.
 At the base level, pairs of halos must obey specific isolation, mass and separation criteria. 
At the second level the orbital angular momentum and energy are constrained and on the third one the phase of the orbit is constrained.
Out of the 300 constrained simulations 146 LGs obey the first set of criteria, 51 the second and 6 the third.
The robustness of our LG `factory'  enables the construction of a large ensemble of simulated LGs. 
Suitable candidates for high resolution hydrodynamical simulations of the LG 
can be drawn from this ensemble, which can be used to perform comprehensive 
studies of the formation of the LG.
\end{abstract}

%%%%%%%%%%%%%%%%%%%%%%%%%%%%%%%%%%%%%%%%%%%%%%%%%%%
%%%%%%%%%%%%%%%%%%%%%%%%%%%%%%%%%%%%%%%%%%%%%%%%%%%

\section{Introduction}\label{sec:intro}

%Cosmology today
These are exciting times for cosmology. Observations of the anisotropies of the cosmic microwave background  (CMB)  radiation by the WMAP and Planck observatories have provided spectacular   validation of the  standard model of cosmology, the $\Lambda$CDM. Observations of distant objects, spanning a look-back time of  over  12 Gyrs, provide further support for the $\Lambda$CDM predictions for the growth of structure in the universe. The basic tenets  of the model  consist of an early inflationary phase, a prolonged phase a homogeneous and isotropic expansion dominated by the dark matter (DM) and dark energy, and structure that emerges out of a primordial perturbation field via gravitational instability.

%Near field cosmology
Cosmology is the science of the biggest possible generalization. It deals with the Universe as a whole.  This leads to an inherent tension between the drive, on the hand, to study the general properties  of everything that we observe and, on the other, the wish to study the particularities of our own patch of the Universe. It follows that cosmology can be practiced by observing the Universe on the largest possible scales. But it can be practiced also by observing the very `local' universe,  resulting in the so-called near field cosmology \citep{Bland-Hawtorn:2006,Bland-Hawtorn:2014}. 
Near field cosmology can test some of the predictions of the $\Lambda$CDM model, and indeed possible conflicts have been uncovered. 
Locally observed dwarf and satellite galaxies seem to be at odd with predictions based on cosmological galaxy formation simulations \citep[e.g.][and references therein]{Peebles:2010,Oman:2015}. 
It is this tension between cosmology, practiced at large, and the near-field cosmology which motivates the Constrained Local UniversE Simulations
(CLUES)  project\footnote{www.clues-project.org} in general and the present paper in particular. Our aim here is to present a numerical laboratory which enables the testing of the near field cosmology against cosmological simulations. 

%Introducing constrained realizations and simulations
The modus operandi of standard cosmological simulations is that they are designed to represent a typical and random realization of the Universe within a given computational box. Indeed, cosmological simulations have been the leading research tool in cosmology and the formation of the large scale structure (LSS). Near field cosmology poses a challenge to the standard cosmological simulations - how to associate environs and objects from the random simulations with  our own Local Group (LG) and its environment?  Bayesian reconstruction methods and constrained simulations provide an alternative to  standard cosmological simulations. The basic idea behind these methods is the use of a Bayesian inference methodology to construct constrained realizations of the local universe of either the present epoch or initial conditions for numerical simulations. 
These constrained realizations are designed to obey a set of observational data and an assumed theoretical prior model. Two main streams have been  followed - one  uses galaxy peculiar velocity surveys and the other galaxy redshift surveys.
\citet{Ganon:1993} were the first to generate constrained initial conditions from peculiar velocity surveys, and these were used by  \citet{Kolatt:1996} to run the first constrained simulations of the local universe. 
This early work has been extended by the CLUES project, within which many velocity constrained simulations have been performed \citep[and references therein]{Gottloeber:2010, Yepes:2014}.  
\citet{Bistolas:1998} and later \citet{Mathis:2002} ran the first redshift survey constrained simulations of the local universe, based on the IRAS survey
\citep{Davis:1991}. 
The application of Bayesian quasi-linear  Hamiltonian Markov Chain Monte Carlo sampling methods to galaxy redshift surveys has provided a new and interesting way to reconstruct the local universe \citep[see e.g.][]{Kitaura:2010, Jasche:2010, Wang:2014, Ata:2015, Lavaux:2016}. 
Yet, these methods have a limited scope of resolution of a few megaparsecs and therefore are unable to resolve the LG itself.    
Presently the only available constrained simulations of the LG are the ones conducted by the CLUES project, to be described below.

% Old CLUES simulations
In the case of the old CLUES simulations \citep[i.e. the ones reported in ][]{Yepes:2014} 
the underlying methodology gives raise to only a small number of  ``realistic`` (numerical) LGs - ones
with mass, distance and relative velocities akin to the observed
LG. In fact, roughly 200 constrained simulations yielded only 4 acceptable LG 
candidates. Though the smallness of the sample hindered  a statistically 
systematic study of the LG within the \LCDM\ cosmology, performing high 
resolution zoom simulations of two of them it has nonetheless been possible to 
address a large number of relevant cosmological and astrophysical issues, such 
as the universality of the DM halo profiles \citep{Mariposa:2013}, properties of substructure \citep{Libeskind:2010},
local implications of the Warm Dark Matter paradigm \citep{Libeskind:2013a} and peculiarities 
of the mass aggregation history of the LG \citep{Forero-Romero:2011}.

One of the major aims of the CLUES project is to turn it into a 'factory' that produces on demand LG-like simulated objects, allowing for a systematic study of the properties of the LG - within the framework of the \LCDM\ model and the Cosmicflows-2 data base of peculiar velocities \citep[CF2;][see the discussion below]{Tully:2013}. The present paper shows  how the incorporation of the CF2 new data and the improved methodology have turned the CLUES project into an efficient 'factory' that produces LGs, essentially on demand. 

% Basic idea behind the CLUES
The basic pillar of the   CLUES approach rests on the fact that  in the standard model of cosmology the primordial perturbation field constitutes a Gaussian random field. The Bayesian linear tools of the Wiener filter (WF) and the constrained realisations (CRs) of Gaussian fields enable the construction 
of ICs constrained both by a given observational data base and an assumed prior model \citep{Hoffman:1991, Zaroubi:1995}.
It follows that there are two attractors that 'pull' the ICs - the prior (cosmological) model and the observational data. Where - either in configuration or in the resolution (k) space - the data is strong the ICs reproduce the constraints, and otherwise they correspond   to random realizations of the prior model.  To the extent that 'strong' data is used as constraints the resulting ICs are likely to reproduce the observed local universe. The so-called constrained variance of the constrained realizations is significantly smaller than the cosmic variance of the prior model. Consequently, the reduction in the cosmic variance measures the 'strength' of the data, for the given prior model.

% The new CLUES simulations
The improvement in the CLUES constrained simulations has proceeded along two main streams. In the methodological one  the original WF/CR algorithm has been amended by the application of the Reverse Zeldovich Approximation (RZA) \citep{Doumler:2013a,Doumler:2013b,Doumler:2013c,Sorce:2014} technique, which accounts for the Zeldovich displacement of the data points. On the data stream, the CF2  \citep{Tully:2013} database of  galaxy peculiar velocities, corrected for the Malmquist bias by the method described by \cite{Sorce:2015}, is used to constrain the ICs. \citet{Sorce:2016} recently presented this new generation CLUES simulations. The present paper extends it to the case of zoom simulations of the LG.

The work starts with \Sec{sec:cs}, that explains the basic ideas behind the Constrained Simulation (CS) method and its numerical implementation.
\Sec{sec:ln} describes the Local Universe, listing the properties which will be used as a benchmark for the quality of the 
reconstruction, while \Sec{sec:simu} discusses the design of the simulations in relation to the aims of the present study.
Then, \Sec{sec:llse} is devoted to the analysis of the reconstructed Local Neighbourhood: The Local Void, the local filament and the Virgo cluster.
\Sec{sec:lg} contains a discussion on the identification and the properties of LG-like pairs, showing that in spite of the large role played by 
the random short-wave modes a substantial number of candidates can be identified 
using distinct classification criteria.
In \Sec{sec:end} the results are summarized and the plan of the future applications of this method is sketched.

%%%%%%%%%%%%%%%%%%%%%%%%%%%%%%%%%%%%%%%%%%%%%%%%%%%%%%%%%%%%%%%%%%%%%%%%%%%%%%%%%%

\section{Constrained Simulations}\label{sec:cs}

At the very core of the constrained simulations lies the \wfcr\ method. 
Its main features will be briefly reviewed in the following subsections, addressing the interested reader to the original works
for a comprehensive theoretical review.

\subsection{Methods}
The WF is a powerful tool for reconstructing a continuous (Gaussian) field from a sparse dataset, assuming a prior model.
Its use in cosmology has been pioneered by \citet{Rybicki:1992} while \citet{Zaroubi:1995} first applied the method to 
the reconstruction of fields from observational dataset.
In practice, the WF reconstruction results in an estimate of the true underlying field which is dominated by the \emph{data} in the region where the
sampling is dense while reproducing the \emph{priors} where the data is lacking or uncertain due to large observational errors.
The recovered velocity field is then easily converted into the cosmic displacement field, which in turn is used to trace back the 
object to its progenitor's position. 
The latter is called the RZA procedure, which shifts backwards the observed \emph{radial} component of the velocity only.
The RZA can be improved replacing the \emph{radial} constraints at these positions by the full three dimensional WF estimation
of the velocity field's components \citep{Sorce:2014}.

However, the recovered velocity field will tend to zero in the regions where data is lacking or dominated by the error, since the WF will tend towards
the mean field, which is the assumed prior. 
This can be compensated by means of the Constrained Realisation algorithm (CR) \citep{Hoffman:1991}, that basically fills
the data-poor regions with a random Gaussian field while converging to the constraints' values where these are present, also
ensuring the overall compliance with the prior's power spectrum.
This means that the outcome of a simulation will be determined by the interplay between these random modes and the constraints.
In particular, the additional random components 
will play a role in two separate regimes: 1) The very large scales, where the data is sparse and error dominated and 2) 
the very small scales, which are dominated by the intrinsic non-linearity of the processes involved and cannot be
constrained by the RZA.
In our case, understanding of the latter kind of randomness is crucial, since it is dominant on the scales that affect
the formation of LG-like objects.

\subsection{Numerical implementation}
The constrained white noise fields are generated using the techniques described by 
\citet{Doumler:2013a, Doumler:2013b, Doumler:2013c, Sorce:2014} on $256^3$ nodes grids, which represent the minimum scale 
on which the constraints are effective \citep{Sorce:2016}.
Short-wave $k$-modes can be added to the input white noise field using the \texttt{Ginnungagap} \footnote{https://github.com/ginnungagapgroup/ginnungagap} 
numerical package.

This numerical setup, that combines two different codes, 
turns out to be extremely relevant to our study, since (as anticipated) one will have to deal with two sources of randomness, i.e.
the one coming from the constrained white noise field, affecting the data-poor large scales, and the small scale one, introduced by 
\texttt{Ginnungagap} when increasing the mass resolution.
This is why, among the other reasons that are explained later, we chose to design and study different sets of simulations that separately relied on the 
two codes, exploiting their different capabilities in different ways and thus allowing us to spot any numerics-induced effect on the results.

%%%%%%%%%%%%%%%%%%%%%%%%%%%%%%%%%%%%%%%%%%%%%%%%%%%%%%%%%%%%%%%%%%%%%%%%%%%%%%%%%%

\section{The Local Universe}\label{sec:ln}

Our definition of the Local Universe encompasses two separate sets of objects; one is
the Local Neighbourhood, which we define as the ensemble of the largest structures 
around the MW, and the other is the Local Group itself. 
Even though our main goal is the simultaneous reproduction of both of them, the differences in scales and methods to be used 
demands a step-by-step separate treatment.
In what follows, a general description of these objects will be provided, with the aim of establishing a benchmark
to gauge the quality of the reconstruction method at different scales and resolutions. 

\subsection{The Local Neighbourhood}
\begin{figure*}
\begin{center}
$
\begin{array}{ccc}
\includegraphics[height=5.5cm]{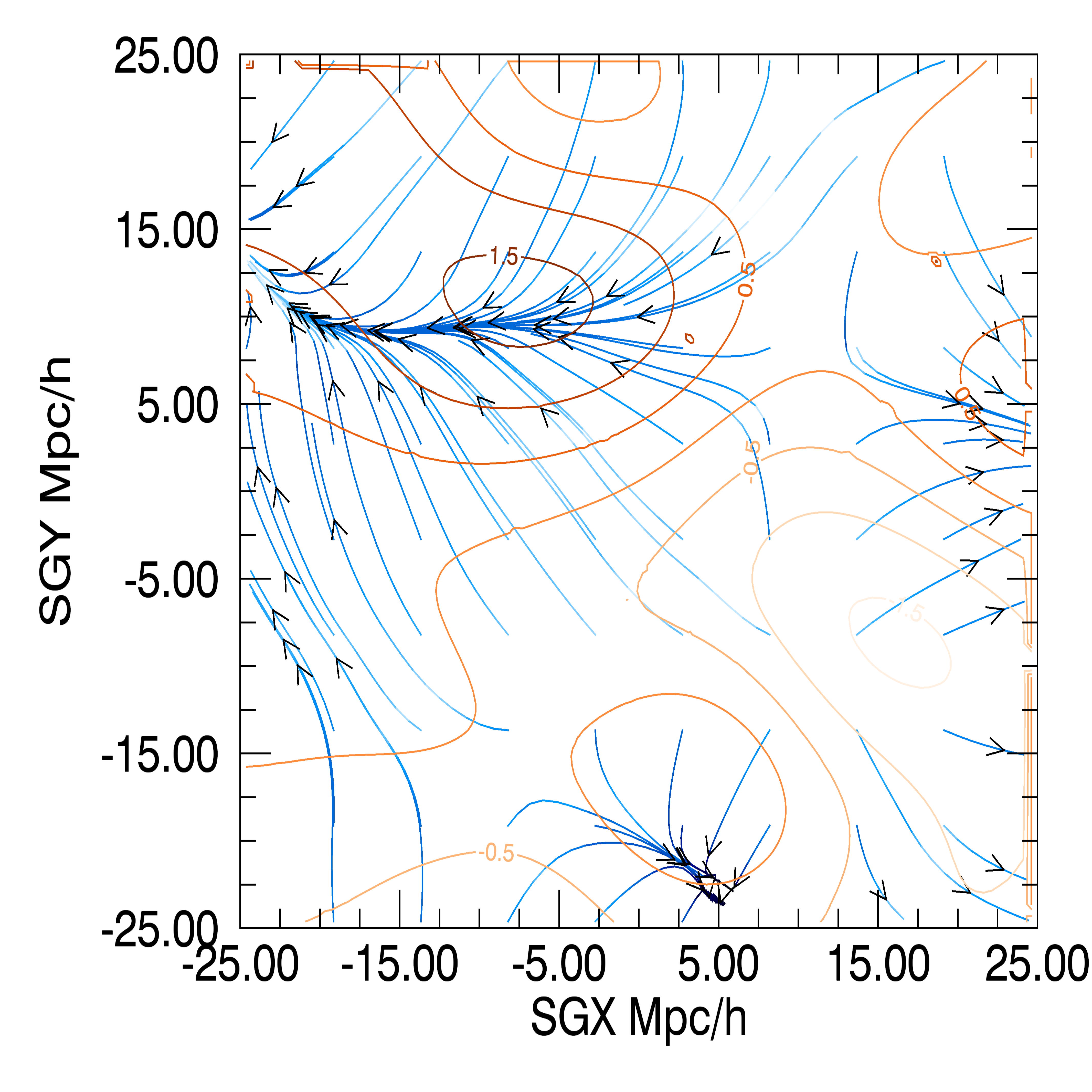} &
\includegraphics[height=5.5cm]{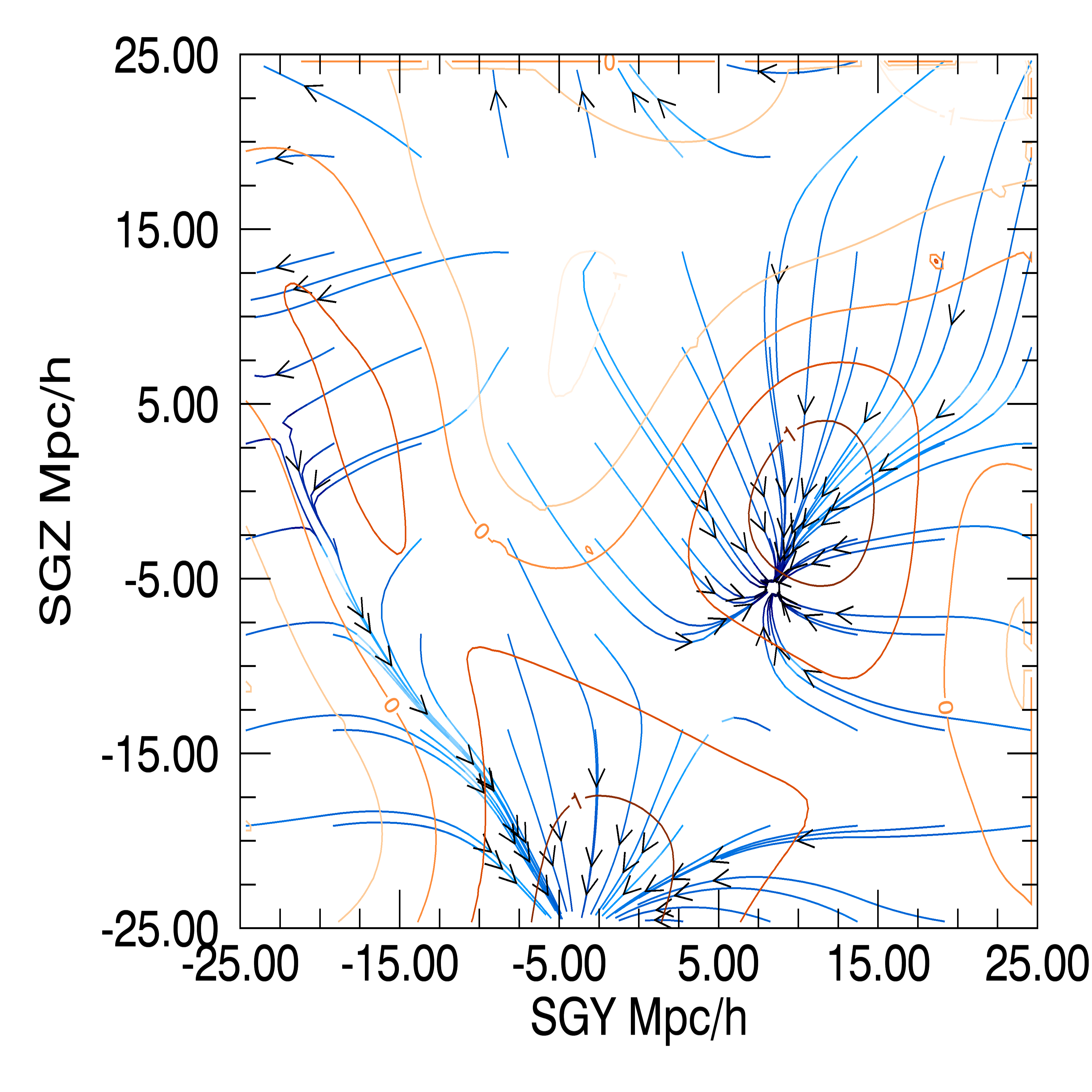} &
\includegraphics[height=5.5cm]{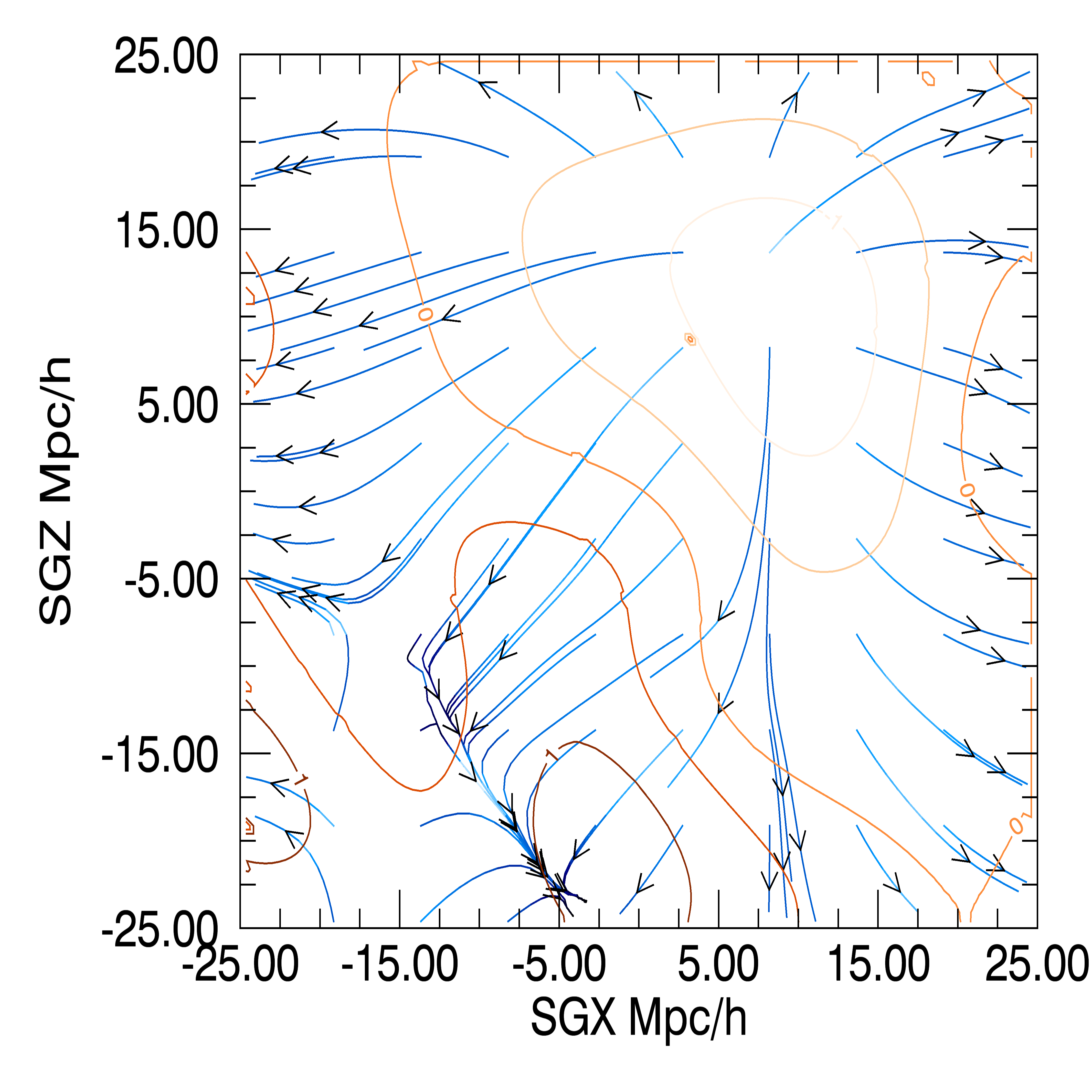} \\
(a) & (b) & (c) \\
\end{array}
$
\caption{
\label{img:wf_cf2}
Linear WF reconstruction of the Local Universe from the CF2 dataset, in the LG frame of reference. 
Velocities and densities are smoothed over $4$\hMpc\ and shown here in
supergalactic coordinates.}
\end{center}
\end{figure*}

Around the supergalactic coordinates $(-3.1, 11.3, -0.58)$\hMpc\ is located Virgo, the closest cluster of galaxies, whose estimated virial mass exceeds
$4\times10^{14}$\hMsun\ \citep{Tully:2009}
and is part of the much larger Laniakea supercluster \citep{Tully:2014}.
Virgo is the most massive object in our neighbourhood, and the effects of its gravitational pull can be observed even at the level of the Local Group, 
since MW and M31 both live in a filament \citep{Klypin:2003a, Libeskind:2015a, Forero-Romero:2015} that stretches from it to the Fornax cluster. 

The main properties of this Local Filament are derived by analysing the eigenvectors of the (observational) velocity shear tensor around the LG position,
which can be recovered from the CF2 dataset \citep{Tully:2013}. 
The CF2 is a catalog of more than 8000 galaxy distances and peculiar velocities; 
which have to be further grouped to smooth out the small scale non-linearities, reducing the final number of constraints to less than 5000 points, 
and treated in order to minimize the biases \citep{Sorce:2015}, 

\Fig{img:wf_cf2} shows the WF reconstruction applied to the CF2 data, 
plotting three density slices of $\pm25$\hMpc\ side and $\pm5$\hMpc\ thickness centered around the 
LG position, showing both the velocity streams and overdensity contours; a similar reconstruction was obtained by \citep{Courtois:2013} using the
CF1 dataset.
This image clearly depicts the nature of the large scale flows in our neighbourhood; for instance, in the supergalactic $X-Y$ plane it can be seen the flow
moving from the center of the box towards Virgo, along the local filament, which then points in 
the direction of the Great Attractor (not shown) towards the negative $X$. 
In the $X-Z$ and $Y-Z$ planes, instead, the effect of the local void can be seen in the $Z>0$ quadrant, 
where the velocity streams clearly indicate an outward movement from such a region.

The local void was first detected and described by \citet{Tully:1987}; however, its nature and extension are still uncertain so 
that estimations of its diameter range from $20$ Mpc \citep{Nasonova:2011} to $> 45$Mpc \citep{Tully:2008}.
It is this large underdense region, situated in the proximity of the LG, which is responsible for the observed 
peculiar velocities of galaxies in its vicinity, including MW and M31 \citep{Tully:2008}.
\citet{Libeskind:2015a} also showed that the center of the local void is aligned with $\hat{\mathbf{e}}_1$ (the eigenvector associated 
with the largest eigenvalue, its numerical value is shown in \Tab{tab:ev_obs})
the direction that defines the strongest infall of matter, again showing the strong influence of this region on the dynamics
in the neighbourhood of the MW.

\begin{table}
\begin{center}
\caption{Eigenvector coordinates and eigenvalues at the LG position for the velocity shear tensor, 
taken from \citet{Libeskind:2015a}. Since eigenvectors are non-directional lines, the $+/-$ directions are arbitrary.
} 
\label{tab:ev_obs}\begin{tabular}{ccccc}
\hline
\quad 			& SGX		& SGY		& SGZ		& $\lambda$ 	\\
\hline
$\hat{\mathbf{e}}_1$	& $-0.331$ 	& $-0.318$	& $0.881$	& $0.148$	\\
$\hat{\mathbf{e}}_2$	& $0.788$ 	& $0.423$	& $0.446$	& $0.051$	\\
$\hat{\mathbf{e}}_3$	& $0.517$	& $-0.848$	& $-0.110$	& $-0.160$	\\
\hline
\end{tabular}
\end{center}
\end{table}

\subsection{The Local Group}

\begin{table}
\begin{center}
\caption{Observational constraints for MW and M31 in the MW frame of reference.
Kinematic properties are taken from \citet{Marel:2012}, distances and errors from with \citet{Marel:2008}, 
while MW and M31 masses are consistent with \citet{Marel:2012, Boylan-Kolchin:2013}. 
}\label{tab:mc_lg}\begin{tabular}{ccc}
\hline
$r_{M31}$      			& kpc	 	 	& $770\pm40$ \\ 
$\mathbf{r}_{M31}$      	& kpc	  		& $(-378.9, 612.7, -283.1)$ \\ 
$\sigma_{\mathbf{r},M31}$       & kpc	  		& $(18.9, 30.6, 14.5)$ \\ 
$\mathbf{v}_{M31}$      	& km s$^{-1}$  		& $(66.1, -76.3, 45.1)$ \\ 
$\sigma_{\mathbf{v},M31}$      	& km s$^{-1}$  		& $(26.7, 19.0, 26.5)$ \\ 
$V_{M31, tan}$ 			& km s$^{-1}$		& $< 34.4$ \\
$V_{M31, rad}$ 			& km s$^{-1}$		& $-109.3\pm4.4$\\
$M_{200, tot}$ 			& $10^{12}$\hMsun   	& $3.14\pm0.58$ \\
$M_{200, MW}$  			& $10^{12}$\hMsun 	& $1.6\pm0.5$ \\
$M_{200, M31}$ 			& $10^{12}$\hMsun	& $1.6\pm0.5$ \\
\hline
\end{tabular}
\end{center}
\end{table}

The term Local Group refers to the group of galaxies dominated by M31 and MW.
There are more than 70 additional galaxies belonging to it, most of whom are of dwarf type and are co-rotating around thin planes centered 
around either of the two hosts \citep{Lynden-Bell:1976, Pawlowski:2013, Ibata:2013}.
The MW and M31 and their satellites form a relatively isolated system, the largest closest object being Cen A, a group of galaxies located at a distance
of around $2.7$\hMpc. 
\Tab{tab:mc_lg} shows the kinematic properties of the LG's main players, 
from which it can be noticed a substantial uncertainty on the mass of the system and a stark
difference between the tangential and radial components of the system's velocity. In particular, the latter property makes the LG
an outlier from the dynamical point of view, since a factor of three difference between these two components in halo pairs of this size
is very rarely observed in cosmological simulations \citep{Fattahi:2015}.

%%%%%%%%%%%%%%%%%%%%%%%%%%%%%%%%%%%%%%%%%%%%%%%%%%%%%%%%%%%%%%%%%%%%%%%%%%%%%%%%%%

\section{The simulations}\label{sec:simu}

The simulation of the Local Universe is a complex task that passes throughout several phases, due to the variety of the
techniques involved and to the different nature (large scale and small scale) of the objects to be simulated.
The present section motivates the choice of the 
design of the simulations in relation to the different aims of the present paper, with a particular
emphasis on the intrinsic limitations of the CS method.

\subsection{Aims}

The properties of the MW and M31 satellites, namely their number and anisotropic distribution, have deep implications for the
\LCDM\ model itself \citep{Klypin:1999b, Kroupa:2005} and is hence one of the strongest motivations for the study of the nearby Universe.
Even though these phenomena cannot be constrained directly, the CS framework provides a powerful and 
robust pipeline to reproduce the
environment in which the LG forms, which is an element that might help explaining the peculiarities of MW and M31 \citep{Gonzalez:2014}.

In the context of simulations, this means that one has to disentangle the randomness of the short-wave modes 
from the large scale constrains and to gauge the effectiveness of CSs and their variance.

Since the reconstruction of the environment is itself a non-trivial task, unaffected by what happens at LG scales,
we decide to take a two steps approach. This way one can ensure that larger structures such as Virgo are well 
recovered without first bothering about what happens to the LG, which is way more problematic due to the predominance of random modes. 
If this is the case, i.e. the reconstructed local neighbourhood compares well to the observational benchmarks previously outlined, one can at least
hope for a successful outcome at smaller scales if some deeper relation between the LG and its environment exists.

In practice, to do this one has to run two series of simulations, addressing different (but related) questions, namely:

\begin{itemize}
\item how stable are the properties of the local neighbourhood? Which role does the random component of the CR plays on these scales?
\item how many LG-like objects can be obtained at the expected position? How can they be properly defined and characterized?
\end{itemize}

In the first case it is clear that there is no need for high resolution, since a sufficiently large number of simulations is enough
in order to address the question of the stability of the results and to estimate the constrained variance of the CS method in the quasi-linear regime.
Even though the CS produces stable results when using boxes of side 500\hMpc, as shown by \citet{Sorce:2016}, 
smaller simulation volumes deserve a separate analysis.
In fact, when dealing with sub-Megaparsec scales it is necessary to deal with the issues of mass resolution 
(which needs to be high in order to reliably identify MW and M31-sized haloes) 
and sample variance (which demands to perform a very large number of simulations). This trade-off can be minimized by the use of zoom-in techniques.
In order to do this while answering to the questions above,
we have designed three series of simulations, which will be described in the following subsection.

\subsection{Settings}

\begin{table*}
\begin{center}
\caption{Simulations settings: Number of realizations per simulation type ($N_{simu}$)
, box size ($L_{box}$, in \hMpc\ units), radius of the Lagrangian
region in the ICs ($R_{zoom}$, in \hMpc\ units), 
effective smallest mass resolution achieved ($m_p$, in \hMsun\ units) and starting redshift ($z_{start}$).
$m_p$ refers to the mass of the particles in the Lagrangian region of the zoom-in \SimuLG\ series; for \SimuLN\ and \SimuLNHi\
it is the mass of all DM particles since they are full box simulations with no higher resolution region.
}

\label{tab:simu}\begin{tabular}{cccccc}
\hline
Name & $N_{simu}$ & $L_{box}$ & $R_{zoom}$ & $m_{p}$ & $z_{start}$ \\
\hline
\SimuLN & $100$ & 100 & NO  & $5.26\times10^{9}$ & $60$  \\
\SimuLNHi & $12$ & 100 & NO  & $6.57\times10^{8}$ & $80$ \\
\SimuLG   & $300$ & 100 & 12 & $6.57\times10^{8}$ & $80$ \\
\hline
\end{tabular}
\end{center}
\end{table*}

In \Tab{tab:simu} are shown the parameter settings for the Initial Conditions (ICs) of the three series of simulations.
The size of the box has been chosen as a compromise between the need for resolution on sub-Mpc scales and the necessity 
to minimize the effects of periodic boundary conditions on Virgo and the local filament. 
All of the simulations are run using the following cosmological parameters: $\Omega_{m}=0.312$, $\Omega_{\Lambda}=0.688$, $h=0.677$ and $\sigma_8=0.807$ 
which are compatible with the Planck-I results \citep{Planck:2013}. To minimize boundary effects and maximize the power of the constraints, the 
observer is placed at the center of the box: This will be considered our expected LG position, around which candidate pairs will be searched for.

This means that the question about LG-like structures has to be addressed by a different type of simulations. The approach in this case has been
    that of adding a sufficiently large region around the box center with an equivalent resolution of $6.57\times10^8$\hMsun, 
which allow to identify this kind of objects with $\approx 10^3$ Dark Matter (DM) particles. 
This is enough to ensure their stability in the process of increasing the resolution by adding further levels of mass refinement.
The size of this first zoom-in region has been chosen as follows. For each \SimuLN\ run,  
 all the particles within a sphere of $R=5$\hMpc\ around the box center (the expected LG location) at $z=0$ were taken and traced back to their 
original positions in the ICs. We then computed the center of mass of these particles and the radius $R_{zoom}$ 
of the smallest sphere enclosing all of them.
It turns out that both these quantities are extremely stable so that a single sphere of $R_{zoom}=10$\hMpc\ placed around the coordinates 
$\mathbf{X}_{init}=(55, 43, 54)$
(\hMpc\ units) would encircle all the 100 spheres generated for each one of the simulations.
Then, to generate the ICs for the LG-tailored simulations, the resolution was increased within a
sphere of $R_{zoom}=12$\hMpc\ around $\mathbf{X}_{init}$.
This pipeline has been tested using several runs with different levels of refinement, in order to ensure that this choice would 
prevent contamination coming from high mass (low resolution) particles within the region of interest for LG-like objects.
This series of zoom-in simulations is labeled \SimuLG. The large scale white noise field is then drawn from 40 out of the 100 \SimuLN\ runs, 
while the short waves necessary to run the zoom-ins were generated using the same set of 10 random seeds for 20 simulations 
and 5 out of those 10 seeds for the remaining ones, making sure not to double count any of the original simulations.

In addition, we ran another series of full box simulations with $512^3$ particles, labeled \SimuLNHi, 
making use of twelve white noise fields (on a $256^3$ grid) from the \SimuLN\ runs and then increasing the resolution.
With this, it was ensured that (1) the resolution of the \SimuLN\ simulations 
was not affecting the results, i.e. the local neighbourhood results largely overlap and (2) that the properties of the zoom in region of the \SimuLG\ runs are the same
of those \SimuLN\ ran with the same pair of large scale and small scale random seeds.
Once these issues have been settled, ensuring that resolution related issues are under control, no further use of the \SimuLNHi\
runs has been made.

All the simulations were executed using the publicly available Tree-PM $N$-Body code \texttt{GADGET2} \citep{Springel:2005} on the MareNostrum supercluster at the 
Barcelona Supercomputing Center, with individual simulations from the \SimuLN\ set running in $\approx60$ core-hours each and the \SimuLG\ ones 
taking approximately twice as much time.

%%%%%%%%%%%%%%%%%%%%%%%%%%%%%%%%%%%%%%%%%%%%%%%%%%%%%%%%%%%%%%%%%%%%%%%%%%%%%%%%%%

\section{Recovery of the local neighbourhood}\label{sec:llse}
\begin{figure*}
\begin{center}
\label{img:llse}
$
\begin{array}{cc}
\includegraphics[height=7.0cm]{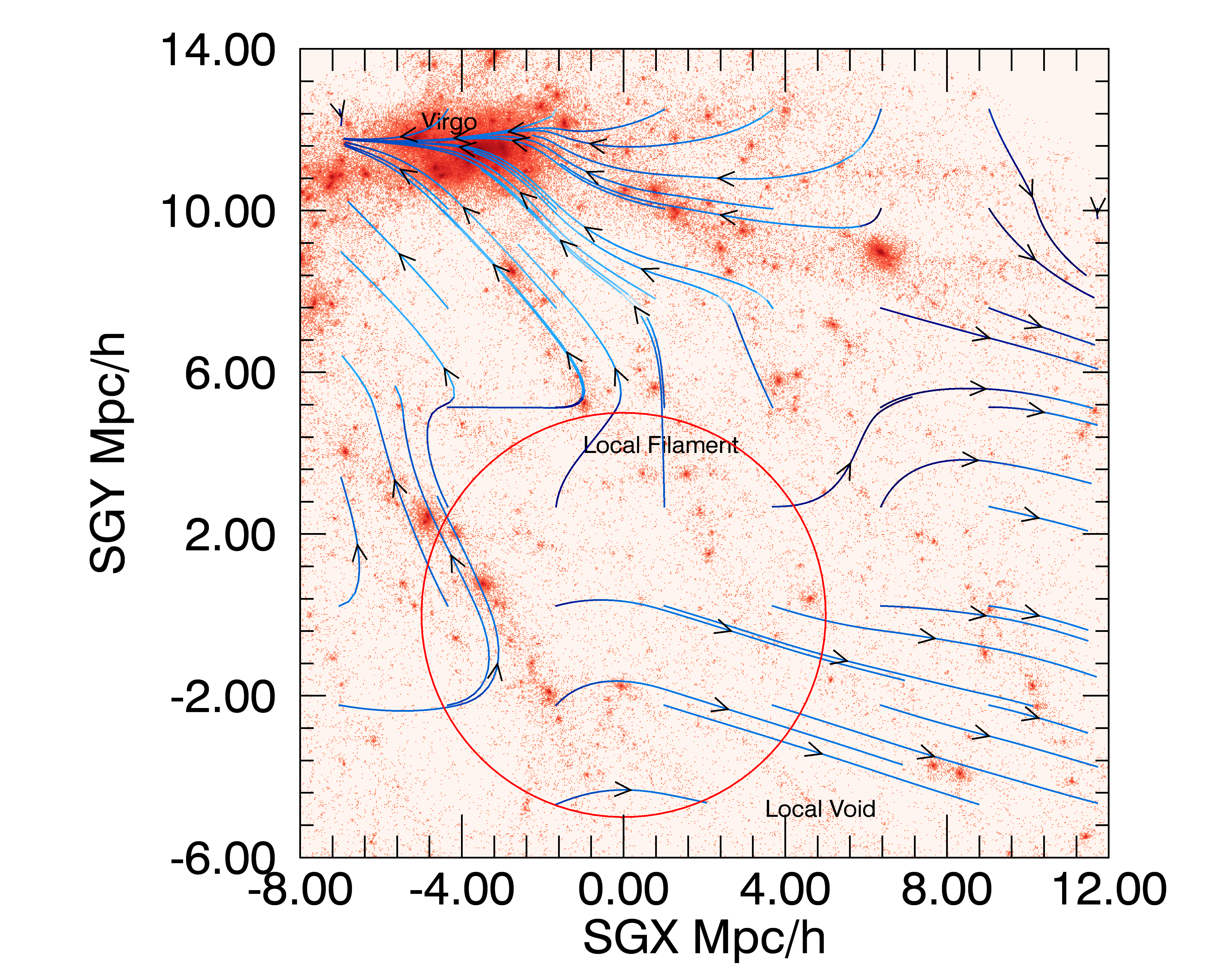} &
\includegraphics[height=7.0cm]{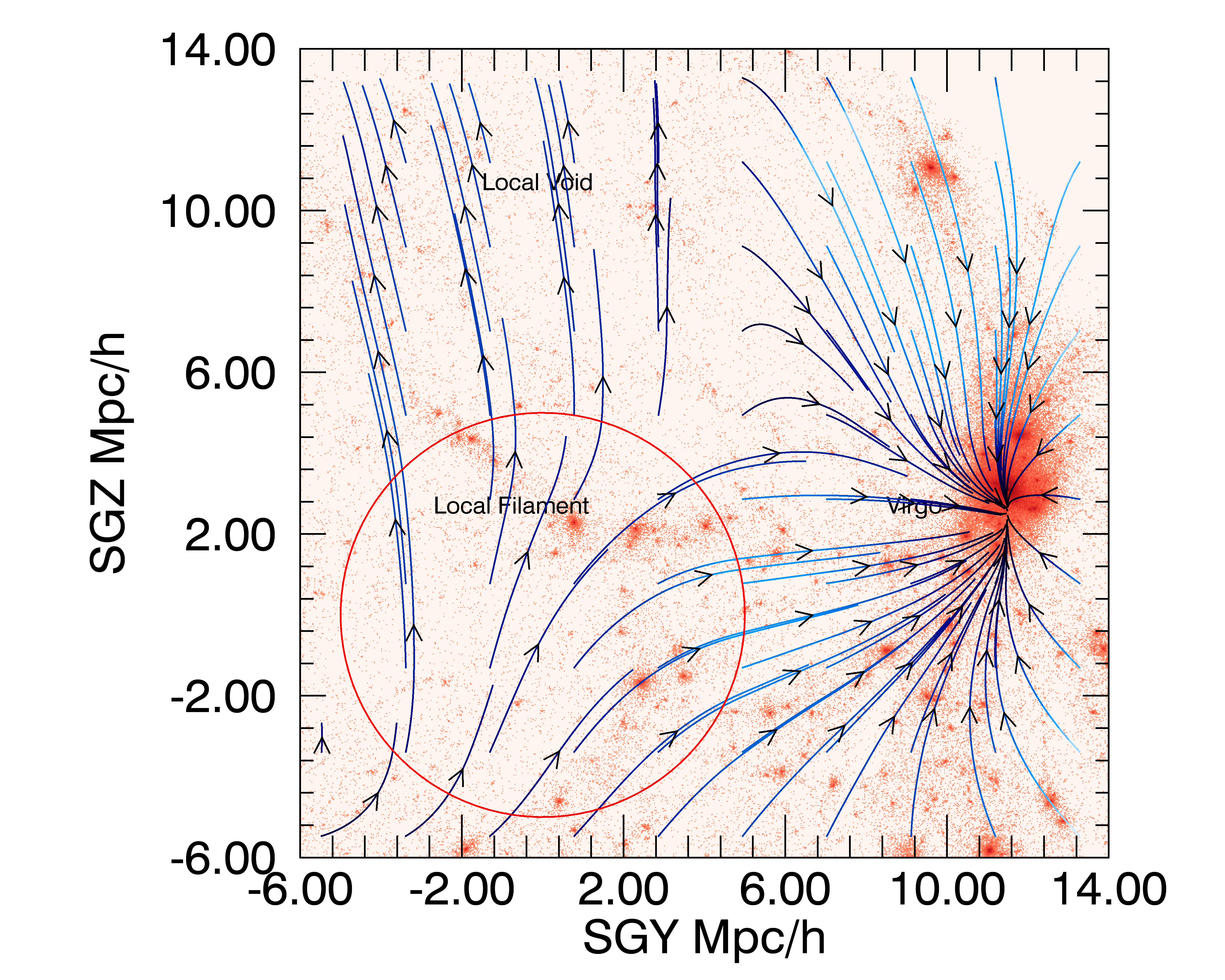} 
\end{array}
$
\end{center}
\begin{center}
\includegraphics[height=7.0cm]{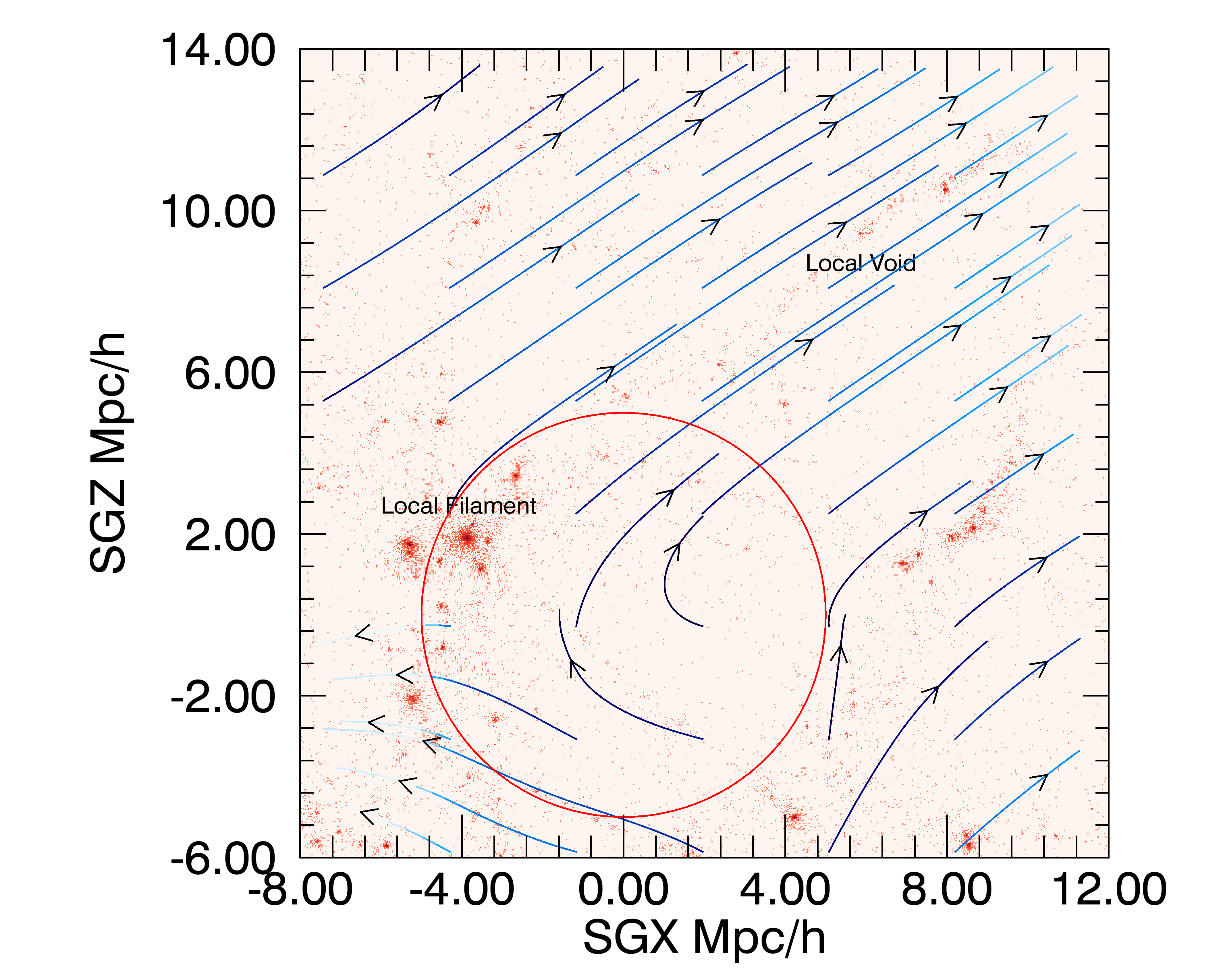} 
\end{center}
\caption{Two dimensional plot of dark matter particles within a $14$\hMpc\ side slice of thickness $\pm3$\hMpc, for the 100 realisations from the \SimuLG\ simulations series, showing the main structures of the local neighbourhood on the $X-Y$, $X-Z$ and $Y-Z$ planes,
in supergalactic coordinates. 
The arrows represent the velocity field as seen in the LG frame of reference.
The red circles enclose the local volume, a sphere of radius 5\hMpc\ that represents the zone of research of LG candidates.
On the $X-Y$ plane it can be seen a thin filament close to the center of the plot and connected to Virgo
on the upper left corner.}
\end{figure*}

This section deals with the issue of the stability of the main large scale structures in the nearby Universe, focusing on the properties of the
Virgo cluster and the local filament. 
The discussion of the local void and its properties will be marginal, since its
reconstruction goes beyond the scope of this paper, due to the smallness of the box size.
A visual example of the reconstruction of the local neighbourhood is presented in \Fig{img:llse}, where all of the aforementioned structures
can be spotted in the dark matter particles distribution and large scale velocity streams.

Naively, one could think that the presence of these objects should be granted by the very nature of the linear WF reconstruction, which is effective
on these scales as shown for the CF2 data reconstruction.
However, the small box and therefore the effect of 
periodic boundary conditions effects make the issue of correct reproduction of the local neighbourhood highly non-trivial, deserving to be treated 
on its own. It is only ensuring that a proper environment stems from the application of the \wfcr\ that one can move forward and look for LG-candidates.

In what follows, the \texttt{AHF} halo finder \citep{Knollmann:2009} has been used to find Virgo and determine its mass and position. 
Different algorithms can be used to define a filament \citep[see e.g.][]{Hahn:2007, Forero-Romero:2009, Sousbie:2011, Falack:2012, Tempel:2013, Chen:2015}, in this work we used
the one described by \citet{Hoffman:2012}, which is based on the dimensionless velocity shear tensor

\begin{equation}\label{eq:shear}
\Sigma_{\alpha\beta} = - \frac{1}{2 H_0} \left ( \frac{\partial v_{\alpha}}{\partial r_{\beta}} - 
\frac{\partial v_{\beta}}{\partial r_{\alpha}} \right)
\end{equation}

\noindent
to classify a region in space given the number of
positive $\Sigma_{\alpha\beta}$ eigenvalues $\lambda$ at that point. Namely, a \emph{void} is defined as an area with no $\lambda>0$, 
a \emph{sheet} has one $\lambda>0$, a \emph{filament} two and a \emph{knot} three. 
The stability of the results has been also checked 
computing the shear tensor over different grids ($128^3$ and $256^3$ nodes) and smoothing lengths (2.5 and 3 \hMpc), showing no particular
dependence on the particular choice of their value \citep{Libeskind:2014}.

In the case of the local void, we will refrain from providing a detailed quantitative analysis of our results, which is left for
future studies.
In fact, due to the limitations imposed by the boundary conditions, we know in advance that it is not 
possible to quantitatively reconstruct a structure of that size within the context of this reconstruction.
It is nonetheless worth noticing that a visual inspection of density contours and velocity streamlines 
consistently reveals the presence of an extended underdense zone in the $Z>0$ area above the center of the box.
Moreover, the shear tensor eigenvectors and eigenvalues show how the push coming from this region is compatible with
the observational estimates of \citep{Libeskind:2015a}, ensuring that our local void reconstruction, albeit incomplete, is
however satisfactory from the point of view of the Local Group and the aims of the current paper.

\subsection{Virgo}
To identify the simulated Virgo candidate, one has to look for the closest and largest structure around the expected
observational position in the \texttt{AHF} catalogs. The properties of these simulated Virgos are listed in \Tab{tab:virgo}.
The first remarkable result is that the variance of the reconstructed masses and positions is small (in agreement with the findings of \citet{Sorce:2016}) 
as can be deducted by looking at both 
the spread between the maximum and minimum values and the standard deviation of each of the properties shown.

\begin{table}
\begin{center}
\caption{Mean and standard deviation of the simulated Virgo mass ($M_{200}$, in $10^{14}$\hMsun) 
and its position in super-galactic coordinates and \hMpc). The corresponding observational values
are taken from the EDD, \citep{Tully:2009}.
All the coordinates are rescaled by shifting the box center from $(50, 50, 50)$ to $(0, 0, 0)$.}
\label{tab:virgo}\begin{tabular}{cccc}
\hline
\quad 		& median & $\sigma$ & $obs.$\\
\hline
$SGX$		& -2.50 & 1.06 & -2.56 \\
$SGY$		& +10.3 & 0.83 & 10.9 \\
$SGZ$		& +1.87 & 1.12 & -0.512 \\
$M_{200}$	& 2.09  & 0.69 & $>4$ \\
\hline
\end{tabular}
\end{center}
\end{table}

This is a proof of the power of this method on these scales, even with a simulation box of only $100$\hMpc, where in principle the effect of the
periodic boundary conditions could be strong enough to spoil the stability of the results.
It is clear that the effect of the random modes is at best marginal 
and the whole reconstruction pipeline results in very stable cluster-sized Virgo look-alikes, 
with a median error in the reconstructed position of only $2.67$\hMpc. As a side note, it can be noticed that the $M_{200}$ is just half of the
observed value for the Virgo mass, which is $> 4\times10^{14}$\hMpc.
This is however an expected result, most likely related to the smallness of the box size. 
In fact, it can be shown \citep{Sorce:2016}
that with a box of $500$\hMpc\ per side the median reconstructed Virgo mass is 
$\approx3.5\times10^{14}$\hMsun, and is thus much closer to the observed value.
Whereas this effect is known for standard $N$-body simulations \citep{Power:2006}, 
a detailed investigation of the correlation between the box size and mass of the local supercluster in CSs 
is beyond the scope of this work and is left for a separate analysis.
For the aim of this paper it is sufficient to prove that our local neighbourhood reconstruction includes a cluster-sized object, placed at the right position,
whose properties are largely independent from the specific random realisation. 
A stable Virgo look-alike is hence the first result of the \wfcr\ pipeline, in agreement with the findings of \citet{Sorce:2016} and a 
substantial improvement over the past outcomes of constrained simulations performed with observational data \citep{Gottloeber:2010},
where reconstructed local superclusters had a position displacement $> 10$\hMpc\ with respect to the expected value. 

\subsection{The Local Filament}
Filaments are identified using the same procedure described in \Sec{sec:llse}, based on the algorithm of \citet{Hoffman:2012}.
It is important to remark that the use of a smoothing length of $2.5$\hMpc\ is dictated by the need to compare to observations, 
filtering out nonlinearities induced by the unconstrained short wave modes. 

The local filament can be found by looking at structures around the center of the box. Since the reconstructed Virgo position turns out to be displaced
by $\approx 2.5$\hMpc\ with respect to the observational value, one would first look for the local filament at a position shifted by an equal amount from the
box center.
However, it turns out that nearest grid point (NGP) of such a displaced position would have (in almost all realisations) only one positive
eigenvalue, meaning that the environment should be classified as a sheet.
Indeed, the center of the box is an excellent choice to look for a filament, since in 
87 out of 100 cases the NGP is characterized by two positive eigenvalues.
Moreover, even in the remaining 13 cases a filament can be always found within $\approx5$\hMpc\ from the center, meaning that this kind
of structure is a constant feature of the expected LG position.

\begin{figure}
\begin{center}
\includegraphics[height=6.0cm]{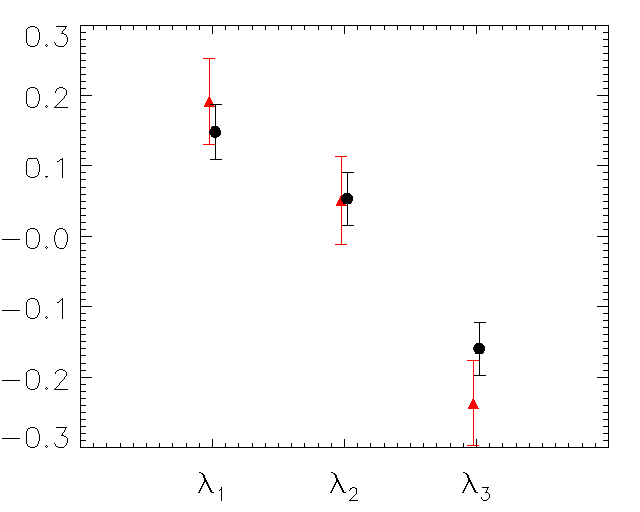}
\caption{Velocity shear tensor eigenvalues and their standard deviations at the center of the box. 
The red triangles are the averages and standard deviations obtained for the
\SimuLN\ while the black circles represent the observational values computed by \citep{Libeskind:2015a}. The two estimates are in very 
good agreement, even though they are computed using different smoothing lengths.
\label{img:ev_wf}}
\end{center}
\end{figure}

\begin{table*}
\begin{center}
\caption{
Comparison between CF2 values for the eigenvectors and eigenvalues at the LG position (taken from \citet{Libeskind:2015a}) and
simulations. Table (a): 
Averages and standard deviations for the sines of the angles of alignment between simulated and observational eigenvectors.
Table (b): Mean values and standard deviations for the eigenvalues at the LG position.
}
\begin{tabular}{cc}

\label{tab:align}\begin{tabular}{cccc}
\hline
$\quad$ & mean & $\sigma$ \\
\hline
$\sin\theta _1$ & $-0.07$ & $0.23$ \\
$\sin\theta _2$ & $-0.19$ & $0.21$ \\
$\sin\theta _3$ & $0.13$ & $0.19$ \\
\hline
\end{tabular} & 

\label{tab:align2}\begin{tabular}{ccc}
\hline
$\quad$ & $simu$ & $obs$ \\
\hline
$\lambda_1$ & $0.174\pm0.062$ & $0.148\pm0.038$ \\
$\lambda_2$ & $0.052\pm0.075$ & $0.051\pm0.039$ \\
$\lambda_3$ & $-0.270\pm0.074$ & $-0.160\pm0.033$ \\
\hline
\end{tabular} \\

(a) & (b) \\

\end{tabular}

\end{center}
\end{table*}

To quantify the goodness of our filament reconstruction we compare our results to \citet{Libeskind:2015a}, which 
were previously shown in \Tab{tab:ev_obs}, 
In \Fig{img:ev_wf} are plotted the $\lambda$s and their standard deviations, 
noticing that our estimated values are in very good agreement with their findings.
This means that the strength of matter accretion (along $\mathbf{e}_1$ and $\mathbf{e}_2$) and expulsion (along $\mathbf{e}_3$) is very well reproduced
by the simulated filament.
Moreover, not only the intensity but also the directions of the eigenvectors are well reproduced, as shown in \Tab{tab:align}.
In fact, the small values of $\sin\theta$ presented there
(defining $\theta$ as the angle between reconstructed $\mathbf{e}_i$ and $\mathbf{e}^{aWF}_i$) indicate a good alignment between the two vectors.
%calculated via the wedge product of the two, 
%Interestingly, it can be noticed that the third eigenvector is the one characterized by the smallest scatter, 
%meaning that the spatial orientation of the filament itself is very well reproduced in our simulations.

%%%%%%%%%%%%%%%%%%%%%%%%%%%%%%%%%%%%%%%%%%%%%%%%%%%%%%%%%%%%%%%%%%%%%%%%%%%%%%%%%

\section{Recovery of the Local Group}\label{sec:lg}

\begin{figure*}
\begin{center}
\label{img:lg}
$
\begin{array}{cc}
\includegraphics[height=7.0cm]{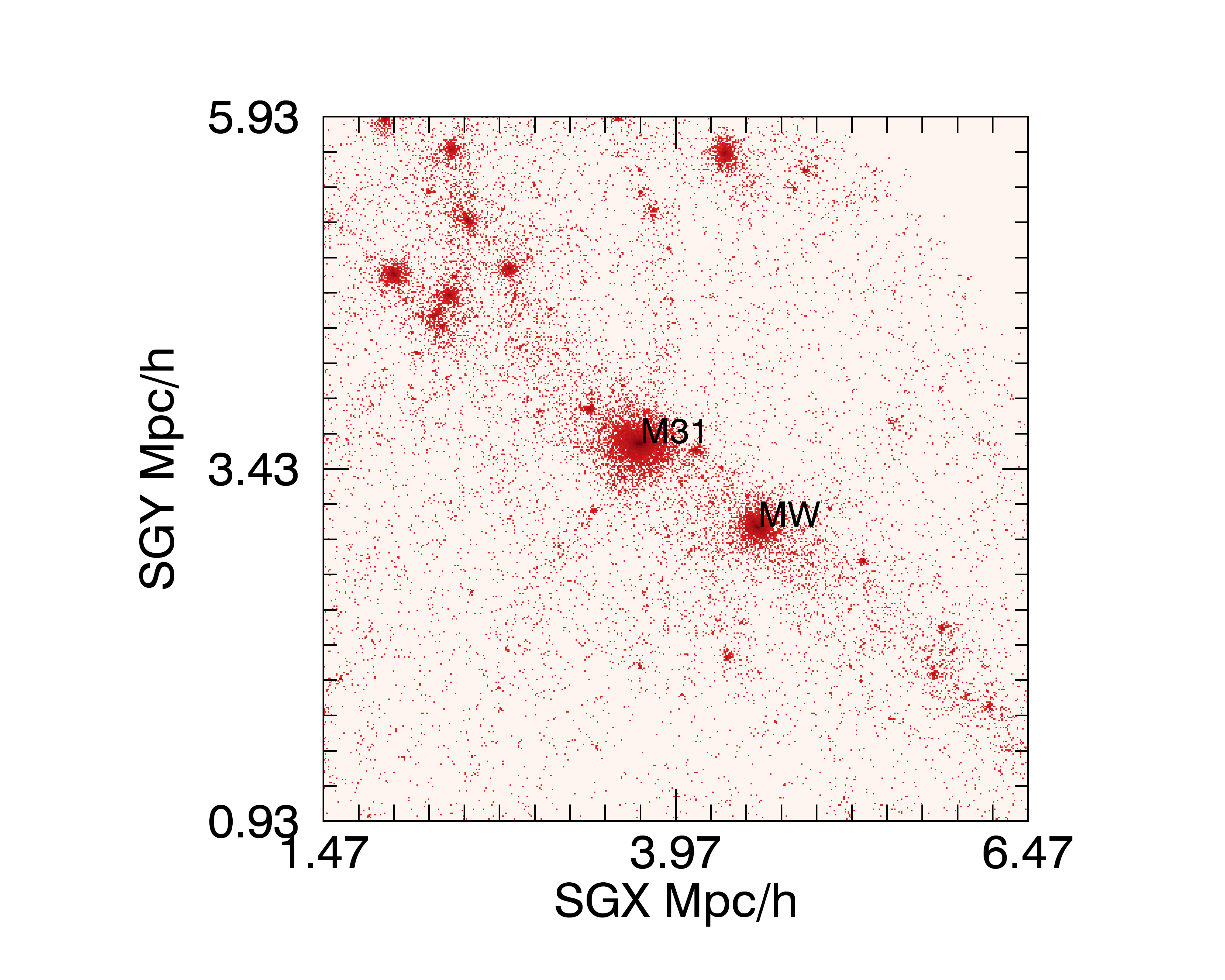} &
\includegraphics[height=7.0cm]{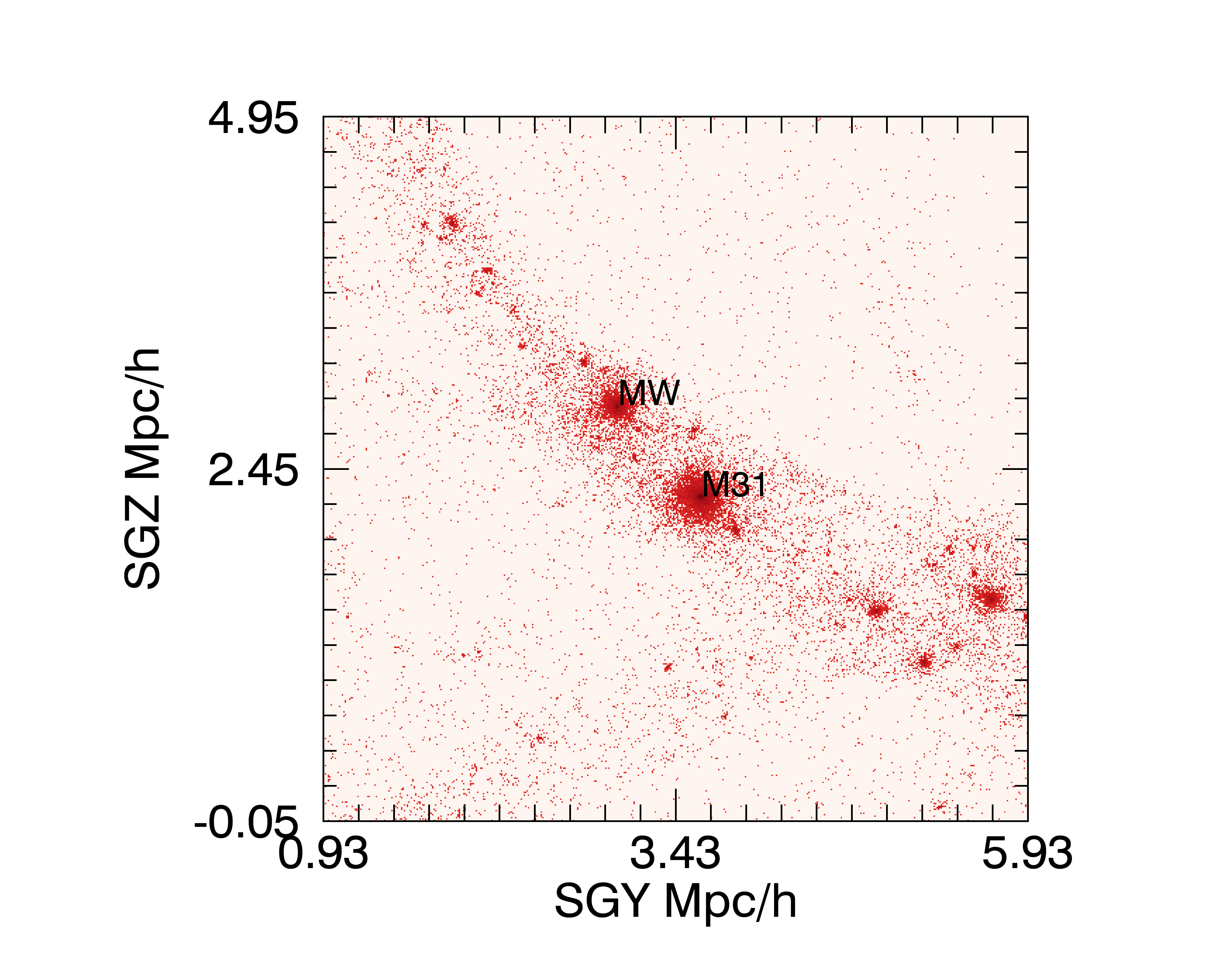} \\
\end{array}
$
\begin{center}
\includegraphics[height=7.0cm]{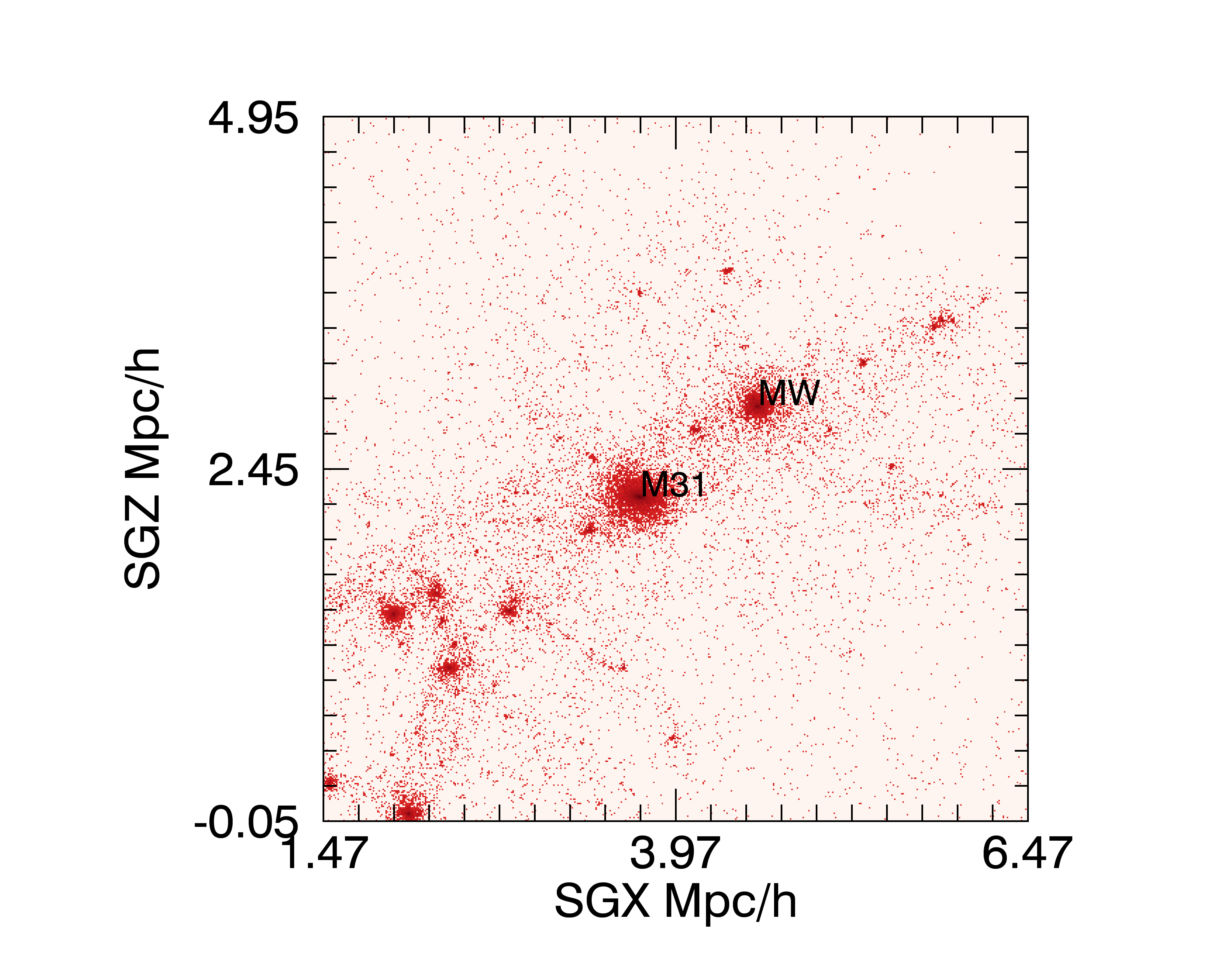} 
\end{center}
\caption{Particle densities in a 5\hMpc\ side slice of thickness $\pm2$\hMpc\ showing the candidates $M31$ and $MW$ in a single realisation, 
along the $X-Y$, $X-Z$ and $Y-Z$ planes. 
}
\end{center}
\end{figure*}

Once it is ensured that the reconstruction methods employed here give rise to a configuration of the local neighbourhood which is both
stable and realistic, one can start to look for possible LG candidates. 
As we discussed in \Sec{sec:simu}, resolution-related problems were bypassed designing a series of zoom-in simulations 
(labeled \SimuLG) with a sufficient number of smaller mass particles just around the expected LG position.
The white noise field of these simulations has been generated in the following way. 
We selected $40$ white noise fields from different SimuLN runs, to have a sufficient
number of different CR of the large scales to factor out possible specific seed-related effects.
Then, for each of these fields were generated 
five to ten different random realisations of the additional layer of refinement in the Lagrangian
region, using \texttt{Ginnungagap} giving rise to different sub-Mpc configurations.
For some of these realisations, it has been checked that the results in the zoom-in regions overlap with those taken from 
the set of \SimuLNHi\ full box $512^3$ simulations.

\subsection{Identification criteria}
The first issue that is encountered in the identification of the LG candidates is related to the details of the definition of a LG-like object.
Indeed, given both the observational uncertainties and uneven relevance of the many LG-defining properties to some general problems, 
the debate on what really constitutes a meaningful candidate is far from settled and several authors in the past have used different
criteria \citep[e.g.][]{Forero-Romero:2013, Gonzalez:2014, Sawala:2014, Libeskind:2015a} to define LG-like pairs in cosmological simulations.
Due to the predominance of the random component at sub-Mpc scales, it is not expected that these objects should be a feature of
the simulations in the same way the local neighbourhood was, so that imposing restrictive identification criteria from the very start might conceal other 
interesting results. 
Moreover, there is a huge number of variables that could be checked and used as criteria to match a simulated LG to the real one.
Therefore, we decided to split the identification issue into several steps, taking into account an increasingly larger subset of variables to
define our candidates.
Namely, we introduced (1) a first broad selection criterion based solely on the
mass and position of isolated halo pairs, (2) a second one based on the requirement that angular momentum and energy
fall within observational bounds and (3) additionally restricting our sample to pairs with 
radial and tangential velocities compatible to the LG values.
In what follows, we will refer to \lgi\ as the first LG-compatibility criterion, \lgii\ as the second one and \lgiii\ as the third one.

Specifically, a \lgi-type LG candidate will satisfy the following requirements:

\begin{itemize}
\item the total mass of the halo pair must be smaller than $5\times10^{12}$\hMsun\
\item the mass of the smallest candidate must be larger than $5\times10^{11}$\hMsun\
\item the distance $d$ between the two haloes must satisfy $0.3 < d < 1.5$\hMpc
\item there is no other halo of mass greater or equal than the smallest candidate within a radius of $2.5$\hMpc
\item the center of mass of the halo pair must be located within 5\hMpc\ from the box center
\end{itemize}

This criterion allows us to obtain a first assessment of the success of our method in producing isolated halo pairs at the right position 
and with a mass of roughly a factor two within the observational constraints on the LG \citep[similar criteria were used by][]
{Forero-Romero:2013, Libeskind:2015a}.
These are general prerequisite characteristics of any realistic LG, and allow us to construct a first large halo sample which could be used to address
a large number of open issues, such as the use of the timing argument \citep{Li:2008, Partridge:2013, Gonzalez:2014}

To identify \lgii\ halo pairs the approach of \citet{Forero-Romero:2013} was followed, considering the (reduced) values of energy angular momentum of the 
\lgi\ candidates to identify those whose \emph{global} dynamic status is compatible with the actual observations.
Defining

\begin{equation}\label{eq:e}
e = \frac{1}{2}\mathbf{v}_{M31}^2 - \frac{GM}{| \mathbf{r}_{M31}|}
\end{equation}

\noindent
as the reduced energy and 

\begin{equation}\label{eq:l}
l = | \mathbf{r}_{M31} \times \mathbf{v}_{M31} |
\end{equation}

\noindent
as the reduced angular momentum, one can 
proceed generating contours in the $e-l$ plane through $10^{7}$ Monte Carlo iterations, realised drawing from the observational values 
listed in \Tab{tab:mc_lg} and assuming Gaussian priors on the $2\sigma$ intervals.
Then, for each \lgi\ pair $e$ and $l$ are computed and all those pairs that fall within the $95\%$ contours are labeled as \lgii-type.
\lgiii\ candidates are those whose radial and tangential velocities fall within $2\sigma$ of the observational values.

\subsection{Local Group like objects}
Using the identification criteria outlined in the previous section, one starts looking for potential LG candidates in each of the 300 \SimuLN\ realisations.
We proceed as follows. First, a sphere of 5\hMpc\ is taken around the box center and all the \lgi\ halo pairs therein are listed.
The environmental type is then determined 
associating each halo to its NGP on a $256^3$ grid, where the eigenvalues of the shear tensor were previously computed.
In this way we can double-check the results of \Sec{sec:llse} and make sure that one can identify
a filament also at the \emph{actual} LG position.
Finally, one can look at the dynamics of the \lgi\ candidates. On the one hand, 
it is interesting to see whether their conserved properties are generally compatible with the observed ones in the sense of the \lgii\ criterion, 
singling out candidate pairs disregarding the specific transient dynamical state.
On the other hand, however, one can be also interested in checking whether some of those pairs actually have radial and tangential velocities in
agreement with observations (type \lgiii). \Tab{tab:lg} shows the number of candidates according to each selection criterion.

\begin{table}
\begin{center}
\caption{Total number of LG-like objects ($N$) and total number of objects located on a filament ($N_f$)
in the \SimuLG\ runs satisfying the \lgi\, \lgii\ and \lgiii\ criteria.}
\label{tab:lg}\begin{tabular}{ccc}
\hline
Type & $N$ & $N_f$\\
\hline
\lgi  & 146 & 120 \\
\lgii & 51  & 42  \\
\lgiii  & 6  & 6  \\
\hline
\end{tabular}
\end{center}
\end{table}

The first important thing to notice is that in approximately half of the simulations (146 out of 300) it is possible to identify a \lgi-type of candidate within
5 \hMpc\ of the expected position. 
These objects are then characterised by the NGP eigenvalues of the two haloes,
allowing the pair to be composed of objects living in different environment types. 
Checking the numbers explicitly for each candidate pair, we find that 120 out of 146 candidates actually 
belong to a filament. 
This is an important results for the CSs: one can produce at a $40\%$ rate objects that can be broadly classified as LGs, within just 
5\hMpc\ from the expected position and that live within a filament.
\Fig{img:lg} offers a visual impression of one of such LG candidates for a single realisation, where the two main haloes can be clearly seen within a filamentary 
stream of dark matter particles.
It is quite clear that such a high success rate has to bear a connection to the shown stability of the local neighbourhood. However, such an investigation is beyond
the scope of the present paper and will be dealt with in a following work. 

One can compare these rates to what has been found by other authors in the literature to quantify the improvement obtained over previous results.
For instance, \citet{Fattahi:2015} found 12 objects looking within a random cosmological simulation box of $100$\hMpc\ per side, 
corresponding to a density of $1.2 \times10^{-5}$\hMpc$^3$. 
Their selection criteria are based on 
\begin{itemize}
\item relative radial velocity between $-250$ and $0$ km s$^{-1}$
\item relative tangential velocity smaller than $100$ 
\item separation between 600 and 1000 kpc
\item total pair mass in the range $\log(M_{tot}/M_{\odot})$
\end{itemize}
Applying these very same constraints on our CSs, a total of 75 objects can be identified.
Since our search volume is made up of 300 spheres of radius $5$\hMpc, the density of LG-like pairs is 
$4.77\times 10^{-4}$\hMpc$^3$, that is a factor of $40$ larger. 
Beside the higher production rate, it has to be stressed that each pair is placed within a large scale environment
that closely matches the observed one, which is the most relevant feature of the CS method.

\begin{figure}
\begin{center}
\includegraphics[height=5.5cm]{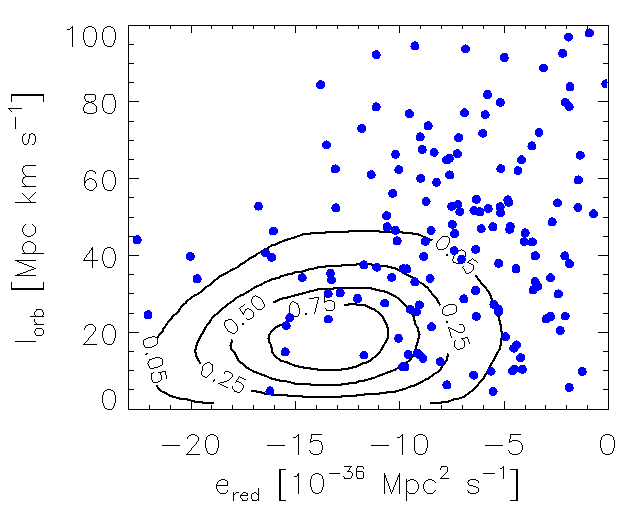}
\caption{
\label{img:candidates}
\lgi\ isolated halo pairs. Solid lines show the dynamical states compatible with LG observations,
determined by $10^7$ Monte Carlo random realisations. Each halo pair that falls within the $95\%$ 
confidence intervals belongs to the \lgii\ sample.}
\end{center}
\end{figure}

To analyze the dynamical properties of this \lgi\ halo sample, the focus is placed on $e$ and $l$, 
the reduced energy of the system and its reduced angular momentum. 
Ideally, in a completely isolated system, these two quantities would be conserved. However, due to the non-ideal nature of the LG this is only partially true. Because of the high degree of isolation implied by our definition of the LG, the expected departure from the ideal case is small. Consequently, it is expected that $e$ and $l$ are almost perfectly conserved, allowing for a more robust classification of the pair.
In fact, it could be misleading to look at distances and velocities only, 
since these quantities are transient by their very nature and might not reveal the fundamental properties of the pair, which might have looked (or will look)
much closer to today's LG values at some other moment in time.
By looking at semi conserved quantities, instead, one is able to single out those structures which have most likely formed with the same local 
initial conditions and compare their histories and evolution. Indeed, the reconstruction in a statistical sense of the history and evolution of the 
LG is one of the ultimate goals of the CLUES collaboration.
As described in the previous paragraph, \lgii\ candidates were identified as those halo pairs falling within the Monte Carlo generated $95\%$
confidence intervals in the $e - l$ plane, selected from the previously identified \lgi-type objects located on a filament.
These candidates are plotted in \Fig{img:candidates}.
From there and from \Tab{tab:lg} one sees that the number of such object is non-negligible, as one can find 42 pairs, corresponding to a $14\%$
success rate. Again, this is a striking result, since it shows that CSs can produce at a significant pace isolated halo pairs whose global
dynamical state is compatible with the observations for the LG. Moreover, these pairs are selected among the larger \lgi\ sample, which 
was already constraining their mass and separation, ensuring that none of these \lgii\ candidates has properties too far off from realistic values.
In fact, it could in principle be possible to produce \lgii-only pairs with unrealistic values of mass and velocities that could 
nonetheless conspire and give rise to a compatible $e-l$ combination.
It has to be noticed that most of the \lgi-candidates lying outside of the $95\%$ confidence intervals are located towards the upper right
corner of the $e - l$ plane, meaning that potential pairs tend to have high angular momenta as well as low binding energies.

As a final test, one can further narrow down the sample to look for \lgiii-pairs, with $v_{rad}$ and $v_{tan}$ within $2\sigma$ from the observational values, 
resulting in a total sample of 6, a factor of ten smaller compared to the total number of \lgii-type objects embedded in a filament. 
This lower number is hardly surprising given the fact that these quantities are transients and are thus even more 
subject to the randomness of the short wave modes. Nonetheless, it is possible to produce a few candidates 
with the right tangential and radial velocities at $z=0$, which could be the starting point for future high resolution zoom in simulations.

\begin{table}
\begin{center}
\caption{Properties of the LG candidates with observationally compatible radial and tangential velocities .
$M_{MW}$ and $M_{M31}$ are expressed in $10^{12}$\hMsun\ units and refer to the smaller and larger
haloes of the pair. $R$ is the distance between the two halo centers (in \hMpc) whereas the tangential and radial velocities are expressed in km s$^{-1}$.
} 
\label{tab:lg3}\begin{tabular}{ccccc}
\hline
$M_{M31}$ & $M_{MW}$ & $R$ & $V_{rad}$ & $V_{tan}$ \\
\hline
  1.54 & 0.59  &    1.45  &   -101.2  &    6.6 \\
  3.08 & 1.11  &    1.01  &   -116.4  &    16.5 \\
  2.49 & 1.19  &    1.32  &   -99.5   &   32.3 \\
  2.08 & 0.61  &    1.48  &   -119.5  &    11.3 \\
  2.80 & 0.97  &    1.03  &   -117.1  &    35.1 \\
  1.01 & 0.73  &   0.59  &   -102.7  &    10.3 \\
\hline
\end{tabular}
\end{center}

\end{table}
 
\subsection{Shear tensor analysis}
\begin{figure}
\begin{center}
\includegraphics[height=5.0cm]{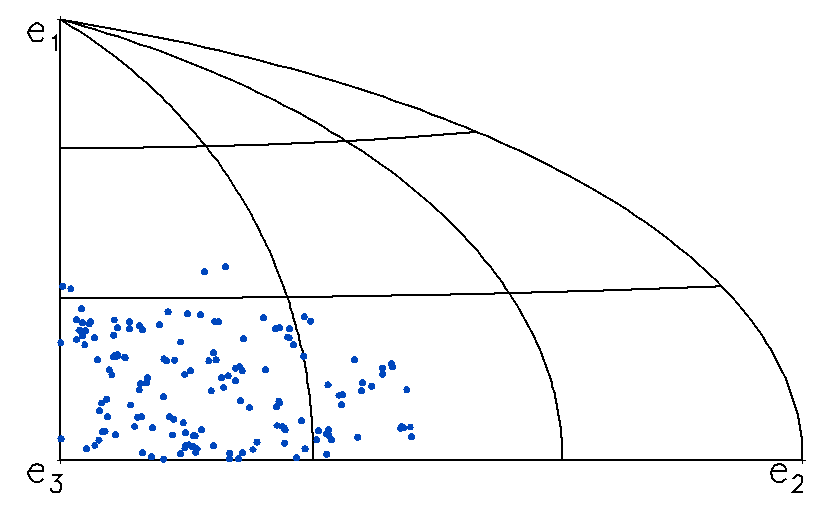}
\caption{
\label{img:lg_lss}
Aitoff projection of Virgo (blue dots) in the eigenvector basis at the LG position for each of the 120 \lgi\ objects
located on a filament, showing that $\mathbf{e}_3$ is pointing towards the direction of Virgo.}
\end{center}
\end{figure}

Having identified a large number of LG candidates one can extend the analysis of the shear tensor eigenvalues and eigenvectors at its position, 
along the lines of what was done in \Sec{sec:llse} at the center of the box, i.e. the \emph{expected} LG position. 
To do this, each halo belonging to the candidate LG is assigned to the NGP, computing the mean and standard deviation of the three $\lambda$s, to 
properly account for those pairs whose members belong to different nodes on the cosmic web. Indeed, as shown in the previous section, only 26 out of 
146 \lgi\ candidates were living on different kinds of environment, whereas the remaining 120 pairs both belonged to a filament, though some of them 
characterized by different sets of eigenvalues and eigenvectors.
Moving to the eigenvector basis in each simulation, we then look for correlations to the large scale environment computing the cosines between 
$\mathbf{e}_1$, $\mathbf{e}_2$, $\mathbf{e}_3$ and $\mathbf{e}_{Virgo}$, the direction to the Virgo cluster. 
From the results shown in \Tab{tab:lg_align} one can see that Virgo is very well aligned with $\mathbf{e}_3$ (which defines the orientation of the filament).
This alignment can be explicitly seen by looking at the Aitoff projection on the quadrant of the sky defined by the three eigenvectors at the $MW$ position, 
in \Fig{img:lg_lss}.
This result is consistent with the findings of \citep{Libeskind:2015a} who found very similar results using the linear WF/CR reconstruction method
on the observational data.

\begin{table}
\begin{center}
\caption{Cosines of the angles between the three eigenvectors at the \lgi positions and 
$\mathbf{e}_{Virgo}$ (the direction of Virgo from the center of mass of the LG), 
showing the alignment of the latter to $\mathbf{e}_3$.}
\label{tab:lg_align}\begin{tabular}{cccc}
\hline
\quad 			& $\mathbf{e}_1$ 	& $\mathbf{e}_2$ 	& $\mathbf{e}_3$ \\  
\hline
$\mathbf{e}_{Virgo}$ 	& $0.28\pm0.24$		& $0.30\pm0.21$		& $0.92\pm0.10$	\\	
\hline
\end{tabular}
\end{center}
\end{table}

To summarize, we have shown that we can use CSs as a \emph{Local Group Factory}: A numerical pipeline for producing 
isolated halo pairs, living on a filament stretched by a large cluster and pushed by a large nearby void, whose properties can be shown 
to be broadly compatible with observational data of the real LG.

%%%%%%%%%%%%%%%%%%%%%%%%%%%%%%%%%%%%%%%%%%%%%%%%%%%%%%%%%%%%%%%%%%%%%%%%%%%%%%%%%%

\section{Conclusions}\label{sec:end}

We have demonstrated 
here how the combined   Cosmicflows-2 data, the \LCDM\  model and the \wfcr\ methodology effectively give rise to Local Group-like objects, in an environment which closely mimics the actual local neighbourhood, in constrained cosmological simulations. The main improvements, in the methodology and the observational data, introduced since the first generation of the CLUES simulations \citep{Yepes:2014} have been reviewed. We rely  heavily on our recent \citet{Sorce:2016} paper and extend it to focus on the simulation of the LG and its close neighbourhood. We  have shown  here how the data, cosmological model and the methodology conspire together to yield a robust 'factory' that produces simulated LG-like objects in abundance.

The simulations reported here are all DM-only and are done within a $100$\hMpc\ box. The \LCDM\ model is assumed with the Planck set of cosmological parameters. Three sets of simulations have been performed: \SimuLN : designed to study the stability of the large scales with respect to the constrained variance; \SimuLNHi : in which the effect of the mass resolution is studied; and \SimuLG :  where zoom-in techniques is used to produce high resolution LGs and are used to study statistically the simulated LGs.

Our main findings fall into two classes. One concerns the environment of the LG and the other deals with the properties of the two halos that constitute the simulated LGs. It should be emphasized  here that essentially all the LGs are embedded in an environment which closely resembles the actual one. A summary of these findings follows:\\
1. The local environment, represented here  by the Virgo cluster and the local filament  are a robust outcome of the 
CF2 data and the \LCDM\ model. Quantitative analysis yields:\\
1a. The mean offset in the position of the simulated Virgo's (from the observed one) is a mere 2.67\hMpc. The median and scatter of the  mass of the simulated Virgo is $2.09 \pm 0.69 \times 10^{14}$\hMsun. This is roughly smaller by a factor 2 from common estimates of the Virgo mass. The smallness of the present computational box accounts for a factor of $\approx 2$ suppression in the mass of the simulated Virgo   \citep[cf.][]{Sorce:2016}.    \\ 
1b. The cosmic web is defined here by the V-web, which is based on the analysis of the velocity shear tensor. The simulated local environments recover the directions of the eigenvectors and magnitude of the eigenvalues.
Roughly 90\% of the simulations recover the local filament at its expected position and 80\% of the identified LGs reside in a filament similar to the observed local filament. The $100$\hMpc\ computational box is too small to reproduce the Local Void, but the analysis of the simulated velocity shear tensor recovers its repulsive 'push' at the position of the LG.\\
2. The frequency, or success rate of the production of simulated LGs: Two sets of criteria have been used to identify LG-like objects. The first consists of the masses of the two main halos, their distance and isolation. The other set adds to the first one also kinematic constraints on semi-conserved quantities, namely the energy and orbital angular momentum. 
The first set of criteria (\lgi) yields isolated halo pairs, neglecting their particular dynamic state, while the second one (\lgii) singles
out candidates whose  energy and angular momentum  are compatible with observations. We have further refined these criteria to objects with the right phase on their orbit, defined by their energy and angular momentum, so as to capture the observed radial and tangential velocities (\lgiii). These sets of criteria yield:\\
2a. Out of a total of 300 \SimuLG\ runs, a total of 146 \lgi-type objects were identified. \\
2b. The fraction of LGs goes down to 17\% - 51 out of 300, when adding energy and orbital angular momentum constraints.
Out of these 51 \lgii candidates only 6
are on the same phase on the orbit  as the actual LG - these are   within $2\sigma$ of the observed 
radial and tangential velocities.  

The LG factory provides a robust tool for producing
ensembles of LG-like objects and their environs. The effort
of obtaining an ensemble of 100, say, LGs, depends on the
definition of a LG. Roughly 200 constrained simulations (of
different realizations of the ICs) are need to reproduce 100
LGs defined by their isolation, mass and separation
distance. Adding the orbital (energy and angular momentum)
constraints the number of required simulations rises to
500. Adding the phase constraint, namely of the radial and
tangential relative velocities, 2700 simulations must be
performed to obtain an ensemble of 100 LGs with the required
properties.

% Main motivation
An inherent tension exists between the fields of cosmology  and the near field cosmology and between the use of random and constrained (local universe) simulations. It relates to the Copernican question of how typical is the LG, and thereby to what extent the LG and the local neighbourhood constitute a typical realization of the universe at large. The answer to this question will determine whether the terms 'near field' and 'cosmology' can be paired together. The LG factory provides a platform for  systematic and statistical studies of the problem and thereby sheds light on the relevance of the near field  to cosmology at large. One qualitative results can already be drawn  from the present study. Our results show that the LG is a likely outcome of the   CF2 data and the assumed \LCDM\ model. 
%Given that the CF2 data itself is a likely outcome of the \LCDM\ model (Hoffman et al. 2016, in preparation), it follows that the LG is a typical realization of the \LCDM\  model.

The present paper serves as a proof of concept for our method. 
It also opens the road to a wide range of applications. 
It enables the construction of a very large ensemble of constrained simulations which harbour realistic LG-like objects and their environment. Such an ensemble can be used to study the (constrained) nature of the LG. Suitable members of the ensemble can be selected as suitable ICs for simulations. Such simulations can include both DM-only and full hydro high resolution simulations. We intend to pursue these and other applications of our method.   

\section*{Acknowledgements}

EC would like to thank the Lady Davis foundation for financial support and Julio Navarro for the fruitful comments and interesting discussions.
JS acknowledges support from the Alexander von Humboldt foundation.
YH has been partially supported by the Israel Science Foundation (1013/12).
SG and YH acknowledge support from DFG under the grant GO563/21-1.
GY thanks MINECO (Spain) for financial support under  project grants AYA2012-31101 and FPA 2012-34694.
SP is supported by the Russian Academy of Sciences program P-41.
AK is supported by the {\it Ministerio de Econom\'ia y Competitividad} (MINECO) in Spain through grant AYA2012-31101 as well as the Consolider-Ingenio 2010 Programme of the {\it Spanish Ministerio de Ciencia e Innovaci\'on} (MICINN) under grant MultiDark CSD2009-00064. He also acknowledges support from the {\it Australian Research Council} (ARC) grants DP130100117 and DP140100198. He further thanks Isaac Hayes for shaft.
We thank the anonymous referee for the useful comments and remarks.
We also thank the Red Espa\~nola de Supercomputaci\' on   for  granting us 
computing time  in the Marenostrum Supercomputer at the BSC-CNS where 
part of the analyses presented in this paper  have been performed.

%%%%%%%%%%%%%%%%%%%%%%%%%%%%%%%%%%%%%%%%%%%%%%%%%%%

\bibliographystyle{mn2e}
\bibliography{biblio}

\bsp

\label{lastpage}

\end{document}